\documentclass{article}
\usepackage{graphicx}
\usepackage{amsmath}
\usepackage{amsfonts}
\usepackage{ dsfont }
\usepackage{ dsfont }

\usepackage{authblk}
\usepackage{braket}

\begin{document}

\title{Quantum Solitons in any Dimension:\\ Derrick's Theorem v. AQFT}
\author[1]{Daniel Davies\thanks{Email: dadavies@ucsc.edu}}
\date{\today}
\affil[1]{\small{Department of Physics, University of California, Santa Cruz, \newline Santa Cruz CA 95064}}

\maketitle

\begin{abstract}
A powerful tool for studying the behavior of classical field theories is Derrick's theorem: one may rule out the existence of localized inhomogeneous stable field configurations (solitons) by inspecting the Hamiltonian and making scaling arguments. For example, the theorem can be used to rule out compact domain wall configurations for the classic $\phi^4$ theory in $3+1$ dimensions and greater. We argue no such obstruction to constructing solitons exists in the framework of algebraic/axiomatic quantum field theory (AQFT), and that states like the example given lie in topologically trivial superselection sectors of the Hilbert space. A proof is presented making use of the relative entropy, and the implications are explored for a few common models of scalar fields and the pure Yang-Mills theory.
\end{abstract}

\newpage
\section{Introduction}

We begin by giving a brief description of Derrick's theorem\cite{1}. It is a result of classical field theory, and may be applied to models containing fields of any spin and of any kind of interaction, so long as a Hamiltonian is specified. In this paper we concern ourselves with only the results as they apply to scalar field theories, for instance the model given by the action of a real scalar field and a non-negative potential:

$$S = \int d^{d+1}x \left(\frac{1}{2}(\partial_\mu \phi)(\partial^\mu \phi) - V(\phi)\right)$$

\noindent The equations of motion for this field are local (by local we mean explicitly that any expression for the field or its derivatives at a point $x$ depends only on the field and its derivatives at the same point $x$ in spacetime), and one may preform a Legendre transform to attain the energy functional. In the case of a time-independent field configuration this functional is

$$H = \int d^dx \left(\frac{1}{2}(\nabla \phi )^2 + V(\phi)\right)$$

Perhaps a localized field configuration $\phi(\vec{x})$ exists with a characteristic scale. We can study the dependence of the energy on this precise scale by varying the coordinates: $\phi(\vec{x})\rightarrow \phi(\lambda \vec{x})$. The Hamiltonian depends on $\lambda$ in the following way:

$$H(\lambda) = \lambda^{d-2} K_2 + \lambda^d U$$

\noindent where $K_2$ and $U$ are the integrated kinetic and potential terms, respectively. For spatial dimension $d \geq 3$, this functional cannot be minimized with respect to $\lambda$, implying the soliton either shrinks or expands indefinitely. This relies on the positivity of $K_2$ and $U$, and that the number of derivatives of the field in the energy functional is less than the number of spacetime dimensions. Indeed, if one adds a 4th derivative term to this functional in $d=3$ dimensions:

$$K_4 = g_4\int d^3x (\nabla^2 \phi)^2$$

\noindent $H(\lambda)$ may be minimized at finite $\lambda$ and classical stable solitons exist in the theory. The constant $g_4$ has dimension $\text{mass}^{-2}$ and so in the renormalized perturbative quantum theory, a model with this term is poorly defined in the ultraviolet. \\

The way solitons are studied in quantum field theory relies heavily on knowledge of the classical solutions to the field equations. The infinite domain wall solution in a $\phi^4$ theory is well studied\cite{2}\cite{3}. The dynamical fields are fluctuations of the fundamental field around a soliton background, and a renormalized perturbation theory is employed to study the model. Examples in other models, e.g. monopoles, cosmic strings, vortices, etc. are all classical solutions to the field equations, and owe their stability in the classical and the quantum theory to conserved topological numbers\cite{2}\cite{3}. The solitons mentioned are all examples of backgrounds which exhibit spontaneously symmetry breaking: a scalar field interpolates between a symmetry-preserving and symmetry-breaking configuration as the distance from the location of the soliton increases. The ``core" of the soliton is a lower-dimensional object where the field is in an exactly symmetric configuration. In the quantum theory, these states exist in a topologically non-trivial sector of the Hilbert space, often specified by a winding number associated with non-trivial homotopy group of the broken symmetry, guaranteeing that no scattering process could change the total topological charge of the Universe. Examples of non-topological solitons are also known\cite{4}, and are referred to as Q-Balls: Q indicating a large Noether charge which serves to stabilize the object. Of particular interest, as we will outline shortly, is the case where solitons are present in strongly coupled theories. An example is the Seiberg-Witten theory\cite{5}, an $\mathcal{N}=2$ supersymmetric Yang-Mills theory with gauge group $SU(2)$. The theory exhibits asymptotic freedom, and at strong coupling, the theory of particles is equivalent to a theory of weakly coupled magnetic monopoles, a consequence of a type of electric-magnetic duality. Furthermore, monopole condensation is the dual description of charge confinement.\\

\noindent We should wonder what it means then for solitons to be present in a theory at strong coupling if no such dual description exists. One might be hopeful and suspect that the lessons of Seiberg and Witten extend to this case as well. The dynamical objects of a theory of particles at strong coupling might generically be weakly coupled solitons. This seems like it might be the case for the strong interactions, though there are difficulties in justifying the effective theory where solitons are manifest. Demonstrating such a thing generically without the use of a duality seems a challenging task, if there is any truth to it. The contrary is of course that the solitons are strongly coupled. If this is the case, the dynamical objects of the theory might only be groups of solitons, though one might wonder if such a theory is physically meaningful in the first place.\\

In this paper we do not approach strong coupling. We study a scenario in which the particles that move in a soliton background interact perturbatively, but we do not approach the subject of calculating perturbative effects in the soliton background. This means that we explore the nature of particle states (both free and bound) that exist near a soliton and move in its influence, but are not self interacting or otherwise. This amounts to an expansion around the soliton background in the action to quadratic order. Local operators are of course constructed using these quantized fluctuation fields, and a perturbation theory may be constructed from the polynomials of the quantized fields. By a perturbative theory, what we mean is one in which asymptotic states may be well approximated by application of a polynomial operator to the corresponding state in the free theory: an element of the Fock space, for instance. This approximation cannot be made arbitrarily close, since these perturbative expansions are generally asymptotic series of a more complicated operator. Before discussing this in more detail, there are two observations we now wish to make about Derrick's theorem that conflict with what is known about quantum theory:\\

1. \textit{The anomalous dimension}. Green's functions in a quantum theory do not obey the scaling relations used above. For example, rescaling of the operator $G_{(2)} = \braket{\partial_\mu \phi(x) \partial^{\mu} \phi(x)}$ is generically of the form $\lambda^{-2+\gamma}G_{(2)}$ with $\gamma$ the anomalous dimension of the operator. $\gamma$ typically depends on the parameters of the model, which in a renormalized perturbation theory, run with the coupling. Indeed, it may be that the anomalous dimension approaches a value of similar order to the classical scaling dimension of the operator (which in the case of Yang-Mills theories occurs when the coupling becomes large\cite{6}), and the scaling argument fails, depending on the sign of the $\gamma$ involved.\\

\noindent Outside of the context of renormalized perturbation theory, quantum theory still does not have trivial scaling relations as in the classical case. Operators only need have support on finite regions of spacetime, and these regions need not be casually connected to the region where an experiment is being preformed to measure the state they correspond to - for example the measuring of the total energy.\\ 

2. \textit{Effective theories are often non-local}. Near and below the $\Lambda_{QCD}$ scale, Green's functions in the local Yang-Mills theory may no longer be computed perturbatively. The degrees of freedom are now described most accurately by the chiral theory\cite{7}, which contains an arbitrarily high number of derivatives. Inclusion of four or more derivatives allows for the existence of topological solitons in the classical theory which are identified with baryons\cite{8}. Exclusion of higher derivative terms is often justified by noting their contributions are suppressed by factors of the pion decay constant, making the calculated scattering amplitudes very accurate when the momentum exchange is low. However, in principle this justification is not satisfying and one would like to compute results when all terms of the chiral Lagrangian are included. Formally, an action which includes an infinite number of derivatives of the fields is non-local. Non-locality of this form is essentially unavoidable in constructing an effective quantum theory in an inhomogeneous background.\\

\noindent Despite the present work taking place in the domain of perturbation theory, one should not lose their conviction that the quantum theory avoids Derrick's theorem at strong coupling; it should instead strengthen. The justification for this is provided in section four.\\

The outline of this paper is as follows. In section 2 we present an argument, based on results of algebraic quantum field theory, that the decay of solitons into final state radiation is prohibited. More specifically, we show that if the domain of support of a particular type of operator is not subject to the Reeh-Schlieder theorem, the vector it corresponds to in the Hilbert space cannot be unitarily evolved into the vacuum sector, i.e. free particle radiation. This implies the existence of a state with non-vanishing overlap with the soliton that is orthogonal to all states of free-particle radiation. To prove such a thing, we demonstrate that the assumption of unitary equivalence is in contradiction with the observed value of the relative entropy for an observer far from the soliton.\\

\noindent In section 3 we explore the implications of section 2 for particular models of interest. First, the 3+1 dimensional $\phi^4$ theory is considered, and we discuss how an effective theory of the solitons must be strictly non-local. Second, we show how to construct single particle states which are bound to the soliton, and we make the case that their contribution to the total energy of the system is what prevents collapse. Third, we make a few comments about the global vortex in 2+1 dimensions. Finally, the free field theory is considered and we show no solitons can be constructed in this case.\\

\noindent In section 4 we briefly discuss the case of strong coupling, and what to expect of solitons in that regime. We argue, drawing examples from QCD, that one should not expect Derrick's theorem to survive in this limit. Finally, in section 5, we summarize our results.\\

\section{A Quantum Derrick's Theorem?}

Before we proceed with the principal argument of this paper, we acknowledge that the algebraic approach to quantum field theory is not taught as a standard graduate course and many readers will be unfamiliar with the relevant theorems or definitions which, for example, are found in the text by Streater and Wightman \cite{9}. To prepare the reader for the remainder of this paper, we now present a brief summary of important definitions and theorems. Revelant proofs are omitted, but briefly described. 

\subsection{Background: The Structure of $\mathcal{H}$, Polynomial Algebras, and the Reeh-Schlieder Theorem}

The Hilbert space $\mathcal{H}$ of a quantum field theory contains a dense subset of vectors $D$ closed under Poincare transformation and a unique state $\Omega \subset D$ (the vacuum) which is Poincare-invariant. The fundamental operators of the theory, the field  $\hat{\phi}_f$ and its adjoint $ \hat{\phi}^*_f$ are tempered distributions, constructed by ``smearing" the physical field $\phi(x)$ with a Schwartz function $f$. Every such smearing function defines a field operator and its adjoint, and the space $D$ is closed under applications of these operators. A smeared field at the point $x$ may be expressed:

$$\hat{\phi}_f(x) = \int d^4y f(x-y)\phi(y)$$

\noindent There is a subset $D_0 \subset D$ spanned by the action of polynomials in the smeared fields (and their adjoints) on $\Omega$, called the \textit{vacuum sector} of $\mathcal{H}$. It is reasonable to restrict the support of these operators (or alternatively, the support of $f$) to an open set $\mathcal{U}\subset \mathcal{M}$ of spacetime. Polynomials formed from operators with support on this open set form the so-called polynomial algebra of the set, $\mathcal{P(U)}$. If two operators are supported on mutually spacelike open sets $\mathcal{U}$ and $\mathcal{V}$, so that every pair of points $x\in\mathcal{U}, y\in \mathcal{V}$ are spacelike separated, then the operators commute.\\

\noindent Naively, one should expect that to create a particle state which is localized on some region of space, the operator which creates the state should be supported on that same region. While this requirement is useful for writing down a ``simple" operator that creates such a particle, it is not strictly necessary.\\

The Reeh-Schlieder theorem\cite{10} states that, essentially, $\mathcal{P(U)}\ket{\Omega} = D_0$. Specifically, every vector in the vacuum sector can be generated by applying a polynomial of fields to the vacuum. The surprising result of the theorem is that there is no restriction on $\mathcal{U}$ other than it be open. The proof of this statement roughly goes as follows: suppose there is an orthogonal state $\chi$ to all those generated by $\mathcal{P(U)}\ket{\Omega}$; the vanishing matrix element is a multilinear functional of the smearing arguments, and so by the Schwartz kernel (nuclear) theorem there exists a unique tempered distribution of all the spacetime coordinates together: 
$$\bra{\chi} \hat{\phi}_f(x_1)\hat{\phi}_f(x_2)\cdots \hat{\phi}_f(x_n) \ket{\Omega} =  W(-x_1, x_1-x_2, \cdots, x_n- x_{n-1}) = 0$$

\noindent The Fourier transform of $W$ vanishes unless each momentum variable lies on the physical spectrum, implying there is a holomorphic function $\mathcal{W}$ of the extended variables $(x_i-x_{i+1})-i\eta_{i+1}$ whose boundary value is $W$, which vanishes for all the $x_i \in \mathcal{U}$. Such a function vanishes for all $x_i$. So $\chi$ is orthogonal to $\mathcal{P(M)}\ket{\Omega} = D_0$, and all states in $D_0$ can be constructed by operators with support on an arbitrary open set. The physical interpretation is this: while it is relatively easy to write down an operator in $\mathcal{P(U)}$ that creates a state localized in $\mathcal{U}$, is is possible (albeit very difficult) to write down an operator in the same algebra that generates a state localized on the other side of the Universe. It can be done, but the polynomial is likely to be of very high order.\\

\noindent There is a corollary to the Reeh-Schlieder theorem that is of some importance. Namely, there is no operator $a\in \mathcal{P(U)}$ where $a\ket{\Omega} = 0$ and $a\neq 0$. This works as long as $\mathcal{U}$ is small enough so that one can find a set which is mutually spacelike from it, $\mathcal{V}$. Suppose $b\in\mathcal{V}$. If the sets are spacelike separated, $a$ and $b$ commute:
$$ba\ket{\Omega} = ab\ket{\Omega} = 0$$

\noindent so that $a$ annihilates (by Reeh-Schlieder) all states in $D_0$. Such an operator must be identically zero.\\

What of $\mathcal{NP(U)}$, the set of operators which is non-polynomial in the smeared fields and has support on $\mathcal{U}$? It is unclear if this forms an algebra, but operators of this type certainly exist, and should send $D_0$ into a larger sector $D_S$. In particular, if there were a conserved (perhaps topological) charge $Q$ in the theory, we might expect that it be defined globally, i.e. the support of $Q$ should contain an entire Cauchy surface, if not all of spacetime. Therefore a state generated uniquely by $Q$ is not in the vacuum sector. We should also be willing to expect that large extended fluctuations of the field also belong to this set, since finite transformations are constructed with the exponential map, which is not polynomial.\\

\subsection{Background:  Cluster Decomposition and The Relative Entropy}

The goal of this section is to show how one might construct a quantum Derrick's theorem, and why such a thing could not work. To do that, we must understand the scaling arguments that can be used in quantum field theory. There are two concepts that are useful here: the cluster decomposition theorem and the relative entropy.\\

The cluster decomposition theorem is a limiting behavior of vacuum expectation values of products of operators. Suppose $a$ has support on $\mathcal{U}$ and $b$ on $\mathcal{U^\prime}$ where $\mathcal{U^\prime}$ is defined by taking all the points of $\mathcal{U}$ and translating them in a spacelike direction $\vec{\lambda}$ so that the two sets are mutually spacelike. As the separation between the sets becomes large, we have 

$$\lim_{\lambda\to\infty} \braket{\Omega|ab|\Omega}= \braket{\Omega|a|\Omega}\braket{\Omega|b|\Omega}$$

\noindent as long as there is a unique vacuum $\Omega$, which we have supposed is true.\\

Before defining the relative entropy, some terminology is in order. A vector $\Psi$ is said to be \textit{cyclic} for an algebra $\mathcal{P(U)}$ if $\mathcal{P(U)}\ket{\Psi}$ is dense in $\mathcal{H}$. It is said to be \textit{separating} if there is no non-vanishing element of the algebra $a$ such that $a\ket{\Psi} = 0$. Clearly the vacuum is cyclic separating for this algebra of interest, provided $\mathcal{U}$ is small enough to be mutually spacelike from another open set. But so are all vectors in $D_0$. For a cyclic separating vector $\Psi$, we can define an anti-linear operator (the relative Tomita operator) $T_{\Psi|\Phi}$ by 

$$T_{\Psi|\Phi} a \ket{\Psi} = a^\dagger \ket{\Phi}$$

\noindent for any $a \in \mathcal{P(U)}$ and any other vector $\Phi$ (it does not need to be cyclic or separating). If $\Psi$ were not cyclic, there would be no sense of a unique adjoint to the relative Tomita operator, and if $\Psi$ were not separating, this operator would not be invertible. Next we define the relative modular operator 

$$\Delta_{\Psi|\Phi} = T^\dagger_{\Psi|\Phi} T_{\Psi|\Phi}$$

\noindent and the relative entropy of the vectors by 

$$S_{\Psi|\Phi}(\mathcal{U}) = -\braket{\Psi|\ln \Delta_{\Psi|\Phi}|\Psi}$$

This construction is a generalization of the Kullback-Leibler divergence to states of quantum field theory, and is a measure of distinguishability between $\Psi$ and $\Phi$. For a more modern and comprehensive review, the reader is encouraged to read reference \cite{11}.
Suppose for a moment that $\ket{\Phi} = \mathbf{u}\ket{\Psi}$, with $\mathbf{u}$ a unitary operator supported on a spacelike separated set $\mathcal{V}$, but also not necessarily an element of $\mathcal{P(V)}$. In this case, $\mathbf{u}$ commutes with elements $a$, so

$$T_{\Psi|\Phi}a\ket{\Psi} = T_{\Psi|\mathbf{u}\Psi}a\ket{\Psi} = a^\dagger \mathbf{u}\ket{\Psi} = \mathbf{u}a^\dagger \ket{\Psi} = \mathbf{u} T_{\Psi|\Psi}a\ket{\Psi}$$

\noindent and so because $\mathbf{u}$ is unitary, one finds that the relative entropy vanishes. The relative entropy has additional traits that make it of importance to us. Firstly, it is non-negative. Secondly, it is monotonically increasing with respect to $\mathcal{U}$. Specifically, if $\mathcal{U} \subset \mathcal{U^\prime}$ then 
$$S_{\Psi|\Phi}(\mathcal{U^\prime}) \geq S_{\Psi|\Phi}(\mathcal{U}) \geq 0$$

\noindent Roughly, a way to understand the monotinicity property is that with a larger set $\mathcal{U}^\prime$ on which to preform local measurements, the more sensitive an observer is to long-range correlations of the field. More information is available and so the uncertainty gained by incorrectly assuming the state of the Universe is $\Phi$ (when it is really $\Psi$) will only increase. This is the classical interpretation of the Kullback-Leibler divergence. Additionally, the only way to saturate the inequality is if the two vectors are related by unitary transformation as described previously. Hence the description of the relative entropy as a measure of distinguishability between $\Psi$ and $\Phi$ is apt.

\subsection{No Obstruction to Solitons}

Suppose a soliton which is not stable in the classical field theory corresponds to a vector $\Psi_S = \mathcal{O}_S\ket{\Omega}$, where $\mathcal{O}_S \in \mathcal{NP(U_S)}$ is supported on a set that contains the physical extent of the soliton. We assert that the Reeh-Schlieder theorem does not apply to this vector: $\mathcal{U}_S$ cannot be chosen arbitrarily and must be large enough to support an algebra that generates single particle states bound to the soliton. To do the latter, it must at least have overlapping support with the region of spacetime that the field is not in a minima of its classical potential.\\

Such a state cannot be unitarily related to a radiation state $\Psi_0 \in D_0$. Assume first that it could be and that the unitary operator which relates them is supported on $\mathcal{U}_S$ or some other set spacelike separated from $\mathcal{V}$. The relative entropy $S_{\Psi_0|\Psi_S}(\mathcal{V})$ is then zero. However, by cluster decomposition, on $\mathcal{V}$ the soliton state cannot be distinguished from the vacuum, or for $a\in \mathcal{P(V)}$:

$$\braket{\Psi_S|a|\Psi_S} = \braket{\Omega|\mathcal{O}^\dagger_S a\mathcal{O}_S|\Omega} = \braket{\Psi_S|\Psi_S}\braket{\Omega|a|\Omega}$$

\noindent (where the operators commute because $\mathcal{O}_S$ necessarily cannot have support on $\mathcal{V}$) which implies
$$S_{\Psi_0|\Psi_S}(\mathcal{V}) \propto S_{\Psi_0|\Omega}(\mathcal{V}) > 0 $$

\noindent Cluster decomposition requires the relative entropy be positive, since far away it is not possible to distinguish the soliton from the vacuum, and it is possible to distinguish the radiation state from the vacuum. However the relative entropy must be zero if the soliton and radiation state are related by the unitary operator we described, so there is a contradiction. What if instead the unitary operator is supported on $\mathcal{V}$, or on any set spacelike separated from $\mathcal{U_S}$? The relative entropy is not automatically zero. However there is still a contradiction, since this would imply $[\mathcal{O}_S,a] \neq 0$ for $a\in \mathcal{P(V)}$, which is a violation of the axiom of local commutativity, given our assumption about the support of $\mathcal{O}_S$.\\

\noindent This argument did not depend on the absolute size of $\mathcal{V}$; we can make it as small as we would like. Suppose there is a Cauchy surface $\Sigma$ that intersects $\mathcal{V}$ and $\mathcal{U}_S$. If $\mathcal{V}$ is made arbitrarily small, its intersection with $\Sigma$ essentially vanishes and the complement becomes all of $\Sigma$. The domain of dependence of this complement, now all of spacetime, is where we had said our unitary operator was supported. Hence the contradiction appears for globally defined unitary operators like the time-evolution operator as well.\\

The contradiction is resolved if we drop the assumption that the vectors are related by unitary transformation (or alternatively, by dropping the assumption that a set $\mathcal{V}$ which is spacelike separated from $\mathcal{U}_S$ even exists, implying that the support of the soliton must extend to an infinite future or past). If a classically unstable field configuration is generated at some initial time by an operator which is not subject to the Reeh-Schlieder theorem, then its vector absolutely must lie in a sector of the Hilbert space orthogonal to the vacuum sector: what we have called $D_S$. It is obvious that no state in $D_S$ can be constructed perturbatively from the vacuum, or from any other state in $D_0$, but it is also true that no unitary time evolution will ever let the two sectors mix.\\

\noindent It remains to be shown that large extended fluctuations fall into this category of operators, or if all non-polynomial operators already satisfy this requirement. We do not know of any operators of this sort that have been written down in the language of the smeared fields, so this requires further investigation. One might expect the operator to look like $\exp\left(i\Delta\phi\hat{\pi}\right)$ in the canonical quantization scheme, which is certainly not a polynomial. In the next section however, we do make a point of constructing single particle states which are bound to the soliton, and demonstrating that their support is not subject to Reeh-Schlieder.\\

\section{Some Examples}

\subsection{Domain Walls in 3+1 Dimensions}

The motivating example for this work was the case of a shrinking compact domain wall, a simple object with an intuitive classical analogue.The transition region from one minima of the classical potential to another produces a shell-like region of high energy density, with an attractive potential $U(r,\theta,\phi)$ and a source $J_0$ localized around the wall (this source is the linear term in the expansion of the Lagrangian density around a soliton background: it vanishes if the soliton's classical equations of motion are satisfied). In the lower dimensional case it is perhaps possible to find a soliton configuration such that $J_0 = 0$, but this is not possible in the higher dimensions, unless the wall is planar and of infinite energy, a consequence of Derrick's theorem. One scenario we can approach analytically is that of a thin spherically symmetric wall of radius $a$, modeled by a delta function:

$$U(r) = m^2 - \frac{3}{2}m^2\delta(r-a)$$
$$J_0 = \frac{c_0m}{ a\sqrt{\lambda}}\delta(r-a)$$

\noindent where $c_0$ is an order 1 constant, determined by some limiting procedure. The dependence on the mass, radius, and quartic coupling $\lambda$ are due to the linear dependence on $\phi_0$ and the mass dimension of $J_0$. The solutions to the Schrodinger equation with $U$ are products of spherical harmonics and spherical Bessel functions, and there is a single eigenvalue $\omega^2$ (for wave functions of the form $f(\vec{x})\cos(\omega t)$), which for large $a$ is approximately $3m^2/4$ and has degeneracy of order $m^2a^2$. For $ma\leq 2/3$, no bound state exists (it has been pushed into the continuous part of the spectrum) and the wall will undergo large fluctuations to approach a stable configuration. For $a$ just larger than this value (but not appreciably, $a \sim 2m^{-1}$ is too large) only an s-wave state is available. For small enough $\lambda$, the mass of these objects is approximated by the classical energy, roughly growing with the surface area and inversely with the coupling (a common trait of solitons): $M_{\text{DW}} \sim  4\pi m^3 r^2 /\lambda$. Another way to express this is via the angular momentum degeneracy of the bound state, $N_\ell$:

$$M_{\text{DW}} \sim  \frac{4\pi m N_\ell}{\lambda}$$

\noindent illustrating that in the weak coupling limit, the mass reasonably scales with the number of quanta that could be trapped on the wall. We suggest, as is the case with the 1+1 dimensional bion, that that presence of these bound states prevents the collapse of the wall.\\

An alternative way to make sense of the failure of Derrick's theorem is to construct an effective action using the path integral formalism. Consider the amplitude for a soliton-to-soliton transition over long times. This defines the path integral in a soliton background, and for simplicity ignore the self-interactions of the fluctuations:
\begin{align*}
\mathcal{Z}_S &= \int \mathcal{D}\eta \exp \left(iS[\phi_0+\eta]\right)\\
&=e^{iS[\phi_0]}\int \mathcal{D}\eta \exp \left(-\frac{i}{2}\braket{\eta|\Delta|\eta} + i\braket{J_0 |\eta} + \text{interactions}\right)
\end{align*}

\noindent where the brackets represent integration over spacetime, in the usual quantum mechanical way. The operator $\Delta$ is a second order linear differential operator and with the function $J_0$, depends explicitly on the background $\phi_0$. If interactions are dropped, the path integral may be solved by direct integration; the result is (symbolically)
$$\mathcal{Z}_S = \exp\left(i S_\text{Eff}[\phi_0]\right)$$
\noindent with
$$S_\text{Eff}[\phi_0] = S[\phi_0] + \frac{1}{2}\text{Tr}\ln\Delta + \frac{1}{2}\braket{J_0|\Delta^{-1}|J_0}$$

\noindent The final term, written out as an integral, is manifestly non-local:

$$\frac{1}{2}\braket{J_0|\Delta^{-1}|J_0} = \frac{1}{2}\iint d^4x d^4y J_0(x)\Delta^{-1}(x,y)J_0(y)$$

\noindent and depends on the $\phi_0$ and its derivatives in a highly non-trivial way. The second (trace) term contributes in a way that depends only on a single spacetime coordinate $x$, but also encodes information about arbitrarily high order derivatives of $\phi_0$. Any equation of motion for the soliton (via collective coordinates or otherwise) will be non-local, obstructing a proof of Derrick's theorem. Alternatively, the exact scaling dimension of $\Delta^{-1}$ or $\ln\Delta$ is not known: a more direct way of obstructing the theorem. Even to all orders in perturbation theory (no longer neglecting interactions), one cannot escape these non-local functionals of the soliton configuration. It is therefore unavoidable that a quantum theory of solitons be non-local, and it is not clear how one would hope to prove Derrick's theorem in this setup.\\

\subsection{Single Particle Bound States}

In the preceding sections we made a point of the presence of bound states on the soliton and the effects they have on its stability. We will now make this point clear, starting with a derivation of the free-particle case borrowed from \cite{12}. In the vacuum sector of a perturbation theory, one constructs single particle states starting with a smeared field operator

$$\hat{\phi}_f(x) = \int d^4y f(x-y)\phi(y)$$

\noindent and restricting the spacetime support of $f$ to any open set $\mathcal{U}$. Additionally, it is necessary to restrict the support of its Fourier transform to the forward mass shell, i.e.

$$\text{supp}(\tilde{f}) = \{k \space|\space m^2 - \epsilon < k^2 < m^2 + \epsilon \}$$

\noindent This guarantees that the state generated by our operator cannot be mistaken for anything but a single particle of mass $m$. Now, to construct the state itself, further smearing with a solution to the Klein-Gordon equation $g(\vec{x},t)$ is needed:

$$\hat{\phi}_{f,g}(\vec{x},t) = -i \int d^3x \left(g(\vec{x},t)\partial_t  \hat{\phi}_f(x)  - \hat{\phi}_f(x) \partial_t g(\vec{x},t)\right)$$

\noindent It is quick to verify that the state this operator creates is time-independent and necessarily has mass $m$:

$$\partial_t \hat{\phi}_{f,g}(t) \ket{\Omega} = -i \int d^3x g(\vec{x},t)\left(\partial^2 + m^2\right)\hat{\phi}_f(x)\ket{\Omega} = 0$$

\noindent where the final equality is true because $\hat{P}^2\ket{\Omega} = 0$. It should be made clear that the spacetime supports of $f$ and $g$ are at least not spacelike separated.\\

How is this construction generalized to the case of the states bound to the soliton? We should expect that their mass is less than $m$, but still part of the discrete spectrum of $\hat{P}^2$. Therefore the initial smeared field is still of use, except its Fourier transform has a different region of support on the forward cone, one which cannot be mistaken for a particle freely propagating far from the soliton. How about the secondary smearing function, $g(\vec{x},t)$? This no longer satisfies the Klein-Gordon equation. Rather, it solves something of the form 

$$\left(\partial^2 + V_0^{\prime\prime}\right)g + J_0 = 0$$

\noindent with $J_0 = \partial^2\phi_0 + V^\prime_0$, $V_0$ the classical potential energy density of the soliton. If the soliton was a completely classical solution, $J_0$ vanishes exactly by the equations of motion. However this is not the case we have been concerned with. Following the procedure for the vacuum sector, we construct $\hat{\phi}_{f,g}(t)$ and find the condition for the state it generates to be time-independent:

$$-i\int d^3x \left(g(\vec{x},t)(\partial^2+V_0^{\prime\prime})+J_0\right)\hat{\phi}_f(x)\ket{\Omega} = 0$$

\noindent Clearly this can be satisfied if $\text{supp}(f,g)$ is far from the soliton, where $J_0 = 0$ and $V_0^{\prime\prime} = m^2$. This is the trivial case. If however we hope to construct states that are localized on the soliton, it should be clear that their support must have significant overlap with $\mathcal{U}_S$. Consider the functions $g$ that actually need to be smeared with $\hat{\phi}_f$ to make this happen. Let's consider the case of a static soliton and construct the zero mode ($\omega^2 = 0$):

$$(-\nabla^2 + V_0^{\prime\prime})g(\vec{x}) + J_0(\vec{x}) = \Delta g + J_0 = 0$$

\noindent This can be further reduced with the ansatz 

$$g = g^\prime - \int d^3y \Delta^{-1}(\vec{x},\vec{y})J_0(\vec{y})$$

\noindent with $g^\prime$ a solution to the zero-frequency Schrodinger equation:

$$\left(-\nabla^2 + V_0^{\prime\prime}\right)g^\prime = 0$$

\noindent This is a state we expect to be localized on the soliton, and to arrive at this conclusion we necessarily required the support of $g$ to overlap with the soliton, specifically the region of the soliton which acts as a source for particle creation, where $J_0$ is non-zero. If a zero mode is not available, the time-dependence of solutions $g(\vec{x},t)$ is not difficult to establish; we need only $0 \leq \omega^2 < m^2$. In summary, to construct a state bound to the soliton, the operator must necessarily have overlapping support with $\mathcal{U}_S$, implying that such a state cannot be subject to Reeh-Schlieder, and therefore cannot be unitarily evolved into a state of free-particle radiation. The underlying reason for this is the non-linearity of the equations of motion for $g$. This implies any expectation value of operators that create the bound states are not multilinear functionals of solutions $g_1, g_2,$ etc. preventing use of the Schwartz kernel theorem. The proof of the Reeh-Schlieder theorem fails.\\

\subsection{The Free Theory}

As a sanity check, we might want to verify that our result is in agreement with what is known about exact solutions to quantum field theories in 3+1 dimensions. Most easily checked is that there are no states in the free field theory other than the ones in the Fock space prescription of the Hilbert space. This is a simple calculation, as the potential for a classical free field (or set of free fields) can be no more complicated than a quadratic form. This implies that $V_0^{\prime\prime} = m^2$ everywhere, no matter the background field configuration. Furthermore, $J_0 = (\partial^2 + m^2)\phi_0$, so the fluctuation field equation is just 

$$\left(\partial^2 + m^2\right)(\eta + \phi_0) = 0$$

\noindent Our ansatz for bound states simplifies exactly to the Klein-Gordon equation

$$(\partial^2 + m^2)\eta^\prime = 0$$

\noindent and no bound states exist. As we have postulated these states serve to stabilize the solitons, and they are not present in the free field theory, it stands to reason that no stable solitons can be formed in the free field theory.\\

\subsection{The Global Vortex}

After the first version of this work was posted, we were made aware of work by Delfino et al. concerning the mass of the global vortex in 2+1 dimensions\cite{13}. The global vortex is a topological soliton in an spontaneously broken $O(2)$ or $U(1)$ model, and has a mass in the classical theory that diverges logarithmically with the volume of space. These authors investigated the vortex solution on a lattice (that is to say, non-perturbatively in the quantum theory) near the critical point and extracted the vortex mass to be $m_V \approx 2.1 m_+$, with $m_+$ the mass of the heavy scalar mode in the trivially broken sector of the theory.\\

\noindent This result, if it is correct, is an explicit example of the violation of Derrick's theorem quantum mechanically, which now would be a statement about there being no\textit{finite energy} inhomogenous stable configuration. In the topologically trivial sector, with a vortex and anti-vortex, Derrick's result indicates a long-range attractive force between the vortices. Classically, one expects the potential energy to depend logarithmically on the separation. If Derrick's theorem is generically violated in quantum theory, we should expect not only that the single vortex has finite mass, but that this long range force is modified appreciably as the vortex anti-vortex separation becomes small. We will address this example in more detail in a forthcoming paper.\\

\section{Remarks on Strong Coupling}

In the introduction we claimed that one's conviction in these results should not waver when strong coupling is introduced. We cannot make a truly convincing argument, as very little can be said about strong coupling except in highly symmetric circumstances. To make our case though, we consider the pure Yang-Mills theory in 3+1 dimensions. The quantum theory is expected to have a mass gap, and exhibit color confinement. Formally, these two properties are somewhat ill-established\cite{14}, but much \textit{is} rigorously known about this theory. For instance, it is asymptotically free and undergoes dimensional transmutation. There is therefore some infrared scale at which strong coupling is unavoidable, and perturbation theory breaks down. The massive dynamical object in this theory is the glueball, and it is thought to be an extended object - a soliton made purely of gauge fields. Lattice approaches to QCD suggest\cite{15} that the glueball is an extended object - a flux tube or string - and that is has significant overlap with a state generated by kind of smeared out Wilson loop\cite{17}. Local polynomial operators in the quantum Yang-Mills theory cannot construct such a thing. Additionally, within the large $N$ limit of $SU(N)$ gauge theory, the glueball mass scales like $N/g^2$, $g$ the dimensionless self coupling \cite{8}. Inverse dependence on a self coupling parameter is a universal feature of known solitons. We will not go into detail here on how our description of a soliton agrees or disagrees with these characterizations, since the present work has focused only on real scalar fields. What we \textit{can} do is illuminate the role Derrick's theorem plays or does not play in classifying the glueball.\\

What should we expect of a classical gauge field configuration that could be identified with a glueball? It should certainly be a localized configuration in space, and probably not identifiable by its bare charge from very far away. On some large spatial surface $S = \partial V$ containing the localized configuration, the color ``Electric" and ``Magnetic" fields should therefore vanish - what is known as \textit{pure gauge} $F_{\mu\nu}^a = 0$. Whether this happens at spatial infinity or at some finite distance from the center of the glueball is of little regard. If we consider a static gauge field $\partial_0 A_\mu^a = 0$, the total energy contained in the volume bounded by $S$ is

$$E = \frac{1}{4g^2}\int d V \text{Tr}\left(F_{\mu\nu}F^{\mu\nu}\right)$$

\noindent in $A_0 = 0$ gauge this simplifies to 

$$E = -\frac{1}{2g^2}\int d V \text{Tr}\left(F_{ij}F^{ij}\right)$$

\noindent Regardless of what gauge we are in, this quantity is bounded from below:

$$E \geq \frac{1}{4g^2}\int d V \text{Tr}\left(F_{\mu\nu}\tilde{F}^{\mu\nu}\right)$$

\noindent If the field is pure gauge everywhere on $V$, the energy vanishes and saturates the inequality. These configurations are the \textit{n-vacua}, so-called because integers label the homotopy classes of maps $S^3 \rightarrow SU(N)$ (identifying the gauge field on the boundary compactifies $V$ into $S^3$). Should $V$ be taken to be all of space, it is not hard to show using the scaling arguments of the introduction that the soliton could decrease its energy by expanding indefinitely. The classical Yang-Mills theory should have no solitons. We can make an even stronger statement by invoking the Chern-Simons form

$$K^\mu = \epsilon^{\mu\nu\lambda\sigma}\left(A^a_\nu F^a_{\sigma\lambda} - \frac{2}{3}f^{abc}A^a_\nu A^b_\sigma A^c_\lambda\right)$$

\noindent where $\partial_\mu K^\mu = \text{Tr}\left(F_{\mu\nu}\tilde{F}^{\mu\nu}\right)$. In the static configuration we have chosen, $\partial_0 K^0 = 0$, so the energy can be made to be bounded from above:

$$E \leq \frac{1}{4}\int d V \partial_i K^i = \frac{1}{4}\int_{S} da^i K^i$$

\noindent where over the domain of integration we enforce a pure-gauge condition. It should not surprise us then (and in fact in static $A_0 = 0$ gauge it is obvious from the definition of $K^i$) that this upper bound on the energy is zero. Take $S$ to be the two-sphere, $S^2$: the homotopy class of maps $S^2 \rightarrow SU(N)$ is trivial, so the integral vanishes in any gauge, not just $A_0=0$.\\

So not only are there no static solitons in the classical pure Yang-Mills theory when all of space is considered, there are also no configurations of positive total energy when a pure gauge condition is enforced on a spatial surface of finite area. Derrick's theorem would not have been so helpful here, since the coordinate transformation introduces a non-trivial dependence of scale into the domain of integration. The classical theory does not give any prescription for constructing soliton-like glueballs, and so the usual approach to studying the quantum mechanics of an extended field configuration - find classical solution and quantize fluctuations around it - fails.\\

\noindent This is precisely why one should not lose confidence that Derrick's theorem cannot be meaningfully extended into field theories with strong coupling. The prototypical example demands solutions to the quantum mechanical field equations that cannot be fundamentally reduced to solutions of the classical field equations, and the only way this can be done is to discard Derrick's theorem completely in one's approach to quantum field theory.

\section{Summary}

We have showed that a result of classical field theory, Derrick's theorem, cannot be extended in a meaningful way into the quantum theory consistent with the axioms as described in \cite{9}. By taking advantage of a self-adjoint operator that encodes information about the algebra of observables, we generalized a scaling argument and were able to show that the proof of Derrick's theorem fails if a particular condition is met. This occurs when a classic result of quantum field theory, the Reeh-Schlieder theorem, fails to describe the physical state of the field and its excitations. We owe the failure of the Reeh-Schlieder theorem (most directly) to the existence of bound single particle states which live on the soliton. There are no unitary operators that interpolate between one of these bound states and an element of the vacuum sector, meaning there is no description of their release into the free-particle spectrum which is consistent with quantum field theory. We believe a description exists in the language of effective action which justifies this: evolution from bound state into free state is kinetically inaccessible. In establishing these statements, we see that the solitons and bound states cannot be in the vacuum sector of the theory. And since no local operator which is a polynomial it the fields can interpolate between the vacuum and the soliton, it must be in a superselection sector of the Hilbert space - one without any topological content.\\

\noindent The reader should not take take this to mean there definitely \textit{are} solitons in any theory which is not identically free. Rather, any no-go theorem constructed in a classical setting is inadequate for making statements about the quantum theory. A systematic approach to finding these solutions must be developed that is consistent with the Wightman axioms. Some conditions have been mentioned in section 3 that serve this purpose, but they are only semi-quantum in nature.\\

The results of this paper apply most directly to scalar fields, and we have not considered the effects of fermions or theories with gauge fields. As we have outlined, the results are somewhat consistent with observed properties of gauge theories, and some version of the statements we made should be expected to hold there. A primary drawback of this work is the use of the polynomial algebra of smeared fields - a construction used to define interacting field theory perturbatively from free fields - to study what we believe to be a fundamentally non-perturbative phenomena. There is undoubtedly an elusive, far richer picture to be seen when sufficient advances are made in non-perturbative quantum field theory. Derrick's theorem however, cannot be an integral part of that picture.\\

\subsection{Acknowledgements}

The author would like to thank his academic advisor, Michael Dine, for the attention and interest Michael expressed in this research these past months. Michael's comments and suggestions helped the author to transform a set of inchoate ideas into the precise and rigorous argument he hopes he has made in these few pages.


\newpage
\begin{thebibliography}
\footnotesize
\bibitem{1}
    G.H. Derrick,
    \textit{Comments on nonlinear wave equations as models for elementary particles },
    J. Math. Phys. \textbf{5} (1964): 1252-1254.
    doi:10.1063/1.1704233
    
\bibitem{2}
    V. Rubakov,
    \textit{Classical Theory of Gauge Fields},
    Princeton University Press, 2002
    
\bibitem{3}
    R. Rajaraman,
    \textit{Solitons and Instantons},
    North-Holland, 1982

\bibitem{4}
    S. Coleman,
    \textit{Q-Balls},
    Nuclear Physics B. \textbf{262} (1985): 263,\\
    doi:10.1016/0550-3213(85)90286-X

\bibitem{5}
    N. Seiberg and E. Witten,
    \textit{Monopole Condensation, And Confinement In N=2 Supersymmetric Yang-Mills Theory},
    arXiv:hep-th/9407087

\bibitem{6}
    M. Peskin and D. Schroeder,
    \textit{An Introduction to Quantum Field Theory},
    Westview Press, 1995; 599-604
    
\bibitem{7}
    G. Ecker,
    \textit{Chiral perturbation theory},
    arXiv:hep-ph/9501357
    
\bibitem{8}
    E. Witten,
    \textit{Baryons in the 1/N Expansion},
    Phys. Lett. B \textbf{86}, 283 (1979), doi:10.1016/0550-3213(79)90232-3
    

\bibitem{9}
    R. F. Streater and A. S. Wightman,
    \textit{PCT, Spin \& Statistics, and All That},
    W. A. Benjamin, 1964

\bibitem{10}
    H. Reeh and S. Schlieder, \textit{Bemerkungen zur Unita¨ar¨aquivalenz von Lorentzinvarienten Feldern,}
    Nuovo Cimento \textbf{22} (1961) 1051

\bibitem{11}
    E. Witten,
    \textit{Notes on Some Entanglement Properties of Quantum Field Theory},
    arXiv:1803.04993
    
    
\bibitem{12}
    A. Duncan, \textit{The Conceptual Framework of Quantum Field Theory},
    Oxford University Press, 2012; 253-271   

\bibitem{13}
    G. Delfino, W. Selke, and A. Squarcini
    \textit{Vortex Mass in the Three-Dimensional O(2) Scalar Theory},
    Phys. Rev. Lett. \textbf{122} (2019), 050602

\bibitem{14}
    A. Jaffe and E. Witten,
    \textit{Quantum Yang-Mills Theory},
    Clay Mathematics Institute, Yang-Mills Existence and Mass Gap; Official Problem Description (available at \textit{claymath.org})
    
\bibitem{15}
    Kenneth G. Wilson,
    \textit{Confinement of Quarks},
    Phys. Rev. D \textbf{10}, 2445 (1974), doi:10.1103/PhysRevD.10.2445

\bibitem{16}
    M. Teper,
    \textit{An improved method for lattice glueball calculations},
    Phys. Lett. B \textbf{183}, 345 (1987),
    doi: 10.1016/0370-2693(87)90976-2

\end{thebibliography}
\end{document}